\documentclass[11pt]{article}
\usepackage{graphicx,hyperref}
\usepackage{
bm}

\newtheorem{prop}{Prop}[section]

\newtheorem{lemma}[prop]{Lemma}

\newtheorem{theorem}[prop]{Theorem}

\newtheorem{notes}[prop]{Note}

\usepackage{amsmath,amssymb}
\usepackage{makeidx}
\makeindex
\usepackage{ascmac}
\usepackage{pb-diagram}
\usepackage{here}
\usepackage{esint}

\title{Gauge theory for mixed $p$-spin glasses}
\author{C. Itoi$^1$  and Y Sakamoto$^2$\\
\\
\vspace{3mm}
$^1$ Department of Physics,   
GS $\&$ CST, 
Nihon University
\\ %
$^2$Laboratory of Physics, CST, 
Nihon University
}

\begin{document}
\maketitle
%
\vspace{10pt}
\begin{abstract}{ 
Physical quantities in the mixed $p$-spin glasses are evaluated with Nishimori's  gauge theory and several variance inequalities.
 The $\mathbb Z_2$-symmetry breaking and the replica-symmetry breaking  are studied 
 in finite and infinite dimensions.
 Obtained bounds on the expectation of the square of the magnetization and spontaneous magnetization
   enable us to clarify  properties of
  paramagnetic and spin glass phases. 
 It is  proven that  variances of ferromagnetic and spin glass
order parameters vanish on the Nishimori line in the infinite volume limit. 
These results imply the self-averaging of these order parameters on the Nishimori line. 
The self-averaging of the 
spin glass  order parameter 
rigorously justifies 
already argued absence of replica-symmetry breaking   on the Nishimori line.
}
\end{abstract}

%
\vspace{2pc}

\section{Introduction} 
 It is well-known that the Parisi formula \cite{Pr0,Pr} indicates the replica-symmetry breaking (RSB)
in the Sherrington-Kirkpatrick (SK) model \cite{SK}.
Rigorous studies on the SK model Guerra started has had great influence on recent spin glass research in mathematical physics
\cite{G,G1}.
Talagrand has developed Guerra's square root interpolation method and
has proven rigorously that the Parisi formula is exact \cite{T2,T}. 
The square root interpolation method developed by Guerra and Talagrand 
is the standard tools for analyzing statistical mechanical models of spin glasses and artificial intelligence \cite{ABBB,AFM,AA,ACCM,BBG,BFT,KXZ,G,G1,T2,T}. 
To understand RSB phenomenon, the self-averaging property and its violation of the spin overlap between different replicas
should be studied. 
The Aizenman-Contucci and Ghirlanda-Guerra (ACGG) identities are useful to study RSB in classical spin systems \cite{AC,CG,CG2,CG3,GG,T}. For example in mean field models, these identities clarify the essential property of spin overlap between different replicas in the Parisi formula in the SK model \cite{Pr0, Pr,T}. 
Toninelli has shown that
the replica symmetric (RS) solution obtained by Sherrington and Kirkpatrick 
becomes unstable \cite{To} in the  region whose boundary is given by the Almeida-Thouless (AT) line \cite{AT}. 
Recently, Chen has proven an important conjecture that the RS solution
becomes exact out of the AT line \cite{C-WK2}.  
Also for short-range interacting models, Chatterjee has proven absence of RSB in the random field Ising model using
 the  ACGG identities \cite{C2}.   
Nishimori's gauge theory is quite useful to understand properties of disordered Ising spin systems 
on the Nishimori line which is a sub-manifold  in the coupling constant space \cite{N0,N}.
The paramagnetic and ferromagnetic phases include the Nishimori line, where Nishimori's gauge theory provides
important informations for several physical quantities, such as  the exact internal energy,  bound on the specific heat  and identities among
correlation functions. On the Nishimori line, a local concavity of the free energy is obtained  by Morita, Nishimori and Contucci \cite{MNC}.
Okuyama and Ohzeki obtained  the Gibbs-Bogoliubov inequality for  the free energy in the SK model using its local concavity 
on the Nishimori line  \cite{OO}.  
These informations are quite helpful to clarify the phase structure of spin glasses in finite and infinite dimensions. 
The fluctuation due to  random interactions  is suppressed on the Nishimori line. The  absence of RSB on the Nishimori line has been argued \cite{NS}. The Nishimori line gives  an important notion
in statistical inference problems and machine learning \cite{KXZ,BDMKZ} in addition to that in spin glasses. 

In the present paper, several identities on the Nishimori line are derived in mixed $p$-spin glass models,  
and utilized them to study spin glasses. 
These identities represent  correlation functions at an arbitrary temperature in terms of  those at the corresponding point on the Nishimori line. 
These representations  enable us to study properties of the spontaneous $\mathbb  Z_2$-symmetry breaking. 
 We show that the sample expectation of the square of the ferromagnetic magnetization 
 and  spontaneous magnetization vanish.
These identities 
and several known inequalities imply that the variances of ferromagnetic and spin glass order parameters  vanish on the Nishimori line  in the infinite volume limit. These results confirm rigorously an already obtained statement that  RSB  does not occur on the Nishimori line \cite{NS}. 

\section{Mixed $p$-spin glasses in finite dimensions}  
\subsection{Definitions}
For a positive integer $L$,  let  $\Lambda_L:= [0,L-1]^d \cap {\mathbb Z}^d
$ be a $d$ dimensional cubic lattice whose volume is $|\Lambda_L|=L^d$. 
 A spin configuration on this lattice is a mapping $\sigma: \Lambda_L \to \{1,-1 \}$  defined by
$
i 
\mapsto \sigma_i =\pm 1$.
Denote a product of  spins 
$$
\sigma_X= \prod_{i \in X} \sigma_i,
$$ 
for a finite sub-lattice $X \subset \Lambda_L$. Let $p$ be  a positive integer. 
To define a short-range $p$-spin Hamiltonian, define a collection ${\cal A}_p$ of
 interaction ranges $A_p \subset \Lambda_L$,  such that
$(0, \cdots, 0) \in  A_p$ and $|A_p| =p$.  
  Define a  collection ${\cal B}_p$ of interaction ranges  by
\begin{equation}
{\cal B}_p:= \{X \subset \Lambda_L | X= i+ A_p, i\in \Lambda_L, A_p\in {\cal A}_p \}.
\end{equation}
Let $Q$ be a finite set of  positive integers.  $Q$ defines 
a Hamiltonian of short-ranged mixed $p$-spin interactions by
\begin{equation}
H(\sigma, \bm J) :=-\sum_{p \in Q} \sum_{X \in {\cal B}_p}  J^p_X
\sigma_{ X},
\label{Hamil}
\end{equation}
where,  a sequence 
$\bm J:=(J_X^p)_{X \in {\cal B}_p , p\in Q}$ consists  of  independent Gaussian 
random variables (r.v.s) with its expectation value  $\mu_p >0$ and its standard deviation $\Delta_p> 0$.
The probability density function of each $J_X^p$ is given by
\begin{equation}
P_p(J_X^p) := \frac{1}{\sqrt{2\pi \Delta_p^2}} \exp \Big[-\frac{(J_X^p-\mu_p)^2}{2 \Delta_p^2}\Big].
\label{PJ}
\end{equation}
${\mathbb E}$ denotes the sample expectation over 
all $J_X^p$, such that  
$$
{\mathbb E} J_X^p=\mu_p, \ \ \ {\mathbb E} (J_X^p -\mu_p)^2=\Delta_p^2.
$$
Gaussian r.v.s $J_X^p$ for 
$X \in {\cal B}_p , p\in Q$ are represented in terms of the independent and  identically distributed (i.i.d.) standard Gaussian
 r.v.s $g_X^p$
\begin{equation}
J_X^p = \Delta_p g_X^p+\mu_p.
\label{stanGauss}
\end{equation}
Note the $\mathbb Z_2$ symmetry
\begin{equation}
H(-\sigma, \bm J) =H(\sigma, \bm J),
\end{equation}
if  all $p \in Q$ are even integers.

\paragraph{Examples of interaction ranges } 
Here, we give several examples of $p$-spin interactions for specific values $p=1,2,4.$
Define a distance 
$|i-j|$ 
between two sites $i:=(i_1,i_2,\cdots, i_d),j:=(j_1,j_2, \cdots, j_d) \in \Lambda_L$
by 
\begin{equation}
|i-j|:=\sum_{s=1}^d|i_s-j_s|.
\end{equation}

\noindent $p=1$.  In this case, the collection of all interaction ranges is given by 
\begin{equation}
{\cal A}_1= \{ 0 \}, \ \ \ {\cal B}_1 = \Lambda_L.
\end{equation}
This corresponds to a random field model.\\

\noindent $p=2$.  In this case, a typical model has exchange interactions of nearest neighbor bonds defined by
\begin{eqnarray}
&&{\cal A}_2:= \{ \{0,i_2\} | \ \  |i_2| =1, i_2 \in \Lambda_L \}, \nonumber  \\ 
&&{\cal B}_2 = \{ \{i_1,i_2 \} | \ \  |i_1-i_2|=1, i_s \in \Lambda_L, s=1,2  \}.
 \end{eqnarray}
The model with nearest neighbor random exchange interactions is the Edwards-Anderson model.\\

\noindent $p=4$.  This case includes plaquette  interactions defined by
\begin{eqnarray}
&&{\cal A}_4:= \{ \{0,i_2, i_3, i_4\} | \ \  1\leq  |i_s| \leq 2, i_s \in \Lambda_L, s=2,3,4 \}, \nonumber  \\ 
 &&{\cal B}_4 := \{ \{i_1,i_2,i_3,i_4\} | \ \ 1\leq |i_s-i_t|\leq 2, i_s\in \Lambda_L, s,t=1,2,3,4  \}.
 \end{eqnarray}
\\

Define Gibbs state for the Hamiltonian.
For a positive $\beta $ and  real numbers $J_X^p$,  the  partition function is defined by
\begin{equation}
Z_L(\beta, \bm J) := {\rm Tr} e^{ - \beta H(\sigma,\bm J)},
\end{equation}
where the trace is taken over all spin configurations. 
Let   $f$ be an arbitrary function 
of spin configuration. The  expectation of $f$ in the Gibbs state is given by
\begin{equation}
\langle f(\sigma) \rangle =\frac{1}{Z_L(\beta, \bm J)}{\rm Tr} f( \sigma)  e^{ - \beta H(\sigma , \bm J)}.
\end{equation}
We define the following functions of  $(\beta,  \bm \Delta , \bm  \mu) \in [0,\infty)^{1+2|Q|} $ and randomness
$\bm J=(J_X^p)_{X \in {\cal B}_p, p \in  Q}$
\begin{equation}
\psi_L(\beta,\bm J) := \frac{1}{|\Lambda_L|} \log Z_L(\beta, \bm J), \\ 
\end{equation}
$-\frac{L^d}{\beta}\psi_L( \beta,\bm J)$ is called free energy in statistical physics.
Define a function $p_L:[0,\infty)^{1+2|Q|} \rightarrow {\mathbb R}$ by
\begin{eqnarray}
p_L(\beta, \bm \Delta, \bm \mu):={\mathbb E} \psi_L(\beta, \bm J ).
\end{eqnarray}
Note that the function $\psi_L(\beta, \bm J)$  and $p_L(\beta, \bm \Delta, \bm \mu)$ 
are convex functions of each variable.

To study ferromagnetic phase transition, 
define an extended  $p$-th ferromagnetic order parameter.
\begin{equation}
m^p := \frac{1}{|{\cal B}_p|} \sum_{X \in {\cal B}_p} \sigma_X
\end{equation}
To study replica-symmetry  and its breaking,  
define $n$ replicated spin configurations $(\sigma^{a})_{a=1,\cdots, n}$
 and a replica symmetric Hamiltonian
\begin{equation}
H(\sigma^1, \cdots, \sigma^n, \bm J):=\sum_{a=1}^n H(\sigma^a, \bm J),
\end{equation}
which  is invariant  under an arbitrary permutation $s \in S_n$  
$$ H(\sigma^1, \cdots, \sigma^n, \bm J)=H(\sigma^{s(1)}, \cdots, \sigma^{s (n) } , \bm J).$$
 We assume replica symmetric boundary condition throughout the present paper.
The covariance of $p$-spin interaction in two replicated Hamiltonians with  indices $a,b \leq n$ 
is defined by the following expectation over the sequence $\bm J^p$ of Gaussian  r.v.s 
\begin{equation} 
 \sum_{X,Y \in {\cal B}_p}{\mathbb E}  J_X^p  \sigma_X^{a} J_Y^p    \sigma_Y^{b} -
\sum_{X,Y \in {\cal B}_p}  {\mathbb E} J_X^p  \sigma_X^{a}   {\mathbb E}  J_Y^p \sigma_Y^{b}
=  |{\cal B}_p| \Delta_p^2 R^p_{a,b},
\end{equation}
where  the $p$-th overlap $R^p_{a,b}$  is defined by
$$
R^p_{a,b}:=\frac{1}{|{\cal B}_p|} \sum_{X \in {\cal B}_p}\sigma^{a}_{X}\sigma^{b}_{X}.
$$
For example,  $p=1$ spin interaction with ${\cal B}_1=\Lambda_L$ is  the random field Zeeman energy, and 
the corresponding  overlap  becomes the site  overlap
$$R^1_{a,b}=\frac{1}{|\Lambda_L|} \sum_{i \in \Lambda_L}\sigma^{a}_i \sigma^{b}_i.$$
The $p=2$ spin interaction with a set ${\cal B}_2 := \{ \{i,j \} | i,j \in \Lambda_L, |i-j|=1 \} $ of nearest neighbor bonds is  the bond exchange interactions, it becomes the bond overlap
$$
R^2_{a,b}=\frac{1}{|{\cal B}_2|} \sum_{ X \in {\cal B}_2}\sigma_X^{a}\sigma^{b}_{X}= 
\frac{1}{ L^d d} \sum_{|i-j|=1} \sigma_i^a  \sigma^a_ j  \sigma^b_i   \sigma^b_j.
$$
In short-range spin glass models, for example  the Edwards-Anderson model \cite{EA} the bond overlap is independent of the site overlap
unlike the SK model \cite{SK}, where the bond overlap is identical to the square of the site overlap. 

Define the Nishimori manifold (NM) by
\begin{equation}
\beta \Delta_p^2 = \mu_p,
\end{equation}
for all $p\in Q$  in the  coupling constant space of $(\beta, \bm \Delta, \bm \mu)
\in [0, \infty) ^{1+2|Q|} $.

\subsection{$\mathbb Z_2$-symmetry breaking}
Let us define a gauge transformation in Nishimori's gauge theory for spin glass \cite{CGN,MNC,N}.
For a spin configuration $\tau \in \{ 1,-1\} ^{ \Lambda_L}$, 
define a gauge transformation by
\begin{equation}
J_X^p \to  J_X^p \tau_X,   \ \ \  \sigma_X \to \sigma_X \tau_X.
\end{equation}
The Hamiltonian is invariant under the gauge transformation.
\begin{equation}
H(\sigma \tau, \bm J \tau ) = H(\sigma, \bm J).
\end{equation}
The distribution function is transformed into 
\begin{equation}
P_p (J_X^p \tau_X)= \tilde P_p (J_X^p) e^{\frac{\mu_p}{\Delta_p^2} J_X^p \tau_X },
\label{gaugetrd}
\end{equation}
where 
$$
 \tilde P_p (J_X^p) :=  \frac{1}{\sqrt{2\pi \Delta_p^2}} \exp \Big[-\frac{(J_X^p)^2+(\mu_p)^2}{2 \Delta_p^2} \Big].
$$
 It is well-known that
 the expectation of the Hamiltonian on NM  is given by
 \begin{equation}
 \mathbb E \langle H \rangle =-  \sum_{p\in Q }|{\cal B}_p|  \mu_p.
 \end{equation}
The following properties of correlation functions on NM are shown in Ref.\cite{ACCM,MNC,N}. 
{\lemma \label{NM} Denote $\beta_{\rm N} := \mu_p/\Delta_p^2$ for all $p\in Q$, then $\beta$ 
is identical to $\beta_{\rm N}$
on the NM.
Denote the Gibbs expectation of an arbitrary function $f: \Sigma_L \to {\mathbb R}$  of spin configuration
$$\langle f(\sigma) \rangle_{\beta}$$
at an inverse temperature $\beta$, if it is necessary.
On NM,  one point function for $X \in {\cal B}_p$ satisfies
\begin{equation}
\mathbb E \langle \sigma_X \rangle _\beta=  \mathbb E \langle \sigma_X  \rangle_\beta \langle \sigma_X  \rangle_{\beta_{\rm N}},
\label{1pointNM}
\end{equation}
and two point  functions for $X,Y \in {\cal B}_p$  satisfy 
\begin{equation}
\mathbb  E \langle \sigma_X \rangle_\beta \langle  \sigma_Y \rangle_\beta =  \mathbb E \langle \sigma_X \rangle_\beta \langle 
\sigma_Y \rangle_\beta \langle \sigma_X \sigma_Y  \rangle_{\beta_{\rm N}},
\ \ \ \mathbb E \langle \sigma_X \sigma_Y \rangle_\beta =  \mathbb E
 \langle \sigma_X \sigma_Y   \rangle_\beta  \langle \sigma_X \sigma_Y   \rangle_{\beta_{\rm N}}. \label{2pointNM}
\end{equation}
An arbitrary multiple point function satisfies an extended formula.}

{\theorem \label{cor}  
Consider the 
model with $Q=\{1,2\}$ for an arbitrary $\beta >0$, and 
define $\beta_{\rm N} := \frac{\mu_2}{\Delta_2^2}$.  The sample expectation of 
square of the ferromagnetic  magnetization vanishes  for any $(\beta, 0, \Delta_2,0, \mu_2)$,
\begin{equation}
\lim_{L\to \infty} \mathbb E   \langle (m^1)^2 \rangle_{\beta} =0,
\end{equation}
if  the quenched expectation of 
square of magnetization vanishes $\displaystyle \lim_{L\to \infty} \mathbb E   \langle (m^1)^2 \rangle_{\beta_{\rm N}} =0,$
for $(\beta_{\rm N}, 0, \Delta_2, 0, \mu_2)$ on NM. \\
 There is no spontaneous magnetization  for any  
 $(\beta, \sqrt{\frac{\mu_1}{\beta_{\rm N}}}, \Delta_2,\mu_1, \mu_2),$
\begin{equation} 
\lim_{\mu_1 \searrow 0} \lim_{L\to \infty} \mathbb E \langle m^1 \rangle_\beta =
\lim_{\mu_1 \searrow 0}\lim_{L\to\infty}\Big[ \frac{1}{\beta} \frac{\partial }{\partial \mu_1} p_L(\beta, \Delta_1, \Delta_2, \mu_1, \mu_2) \Big]_{\Delta_1 = \sqrt{\frac{\mu_1}{\beta_{\rm N}}} }
 =0,
\end{equation}
 if there is no spontaneous magnetization 
 $\displaystyle
 \lim_{\mu_1 \searrow 0} \lim_{L\to \infty} \mathbb E \langle m^1 \rangle_{\beta _{\rm N}}=0$
for $(\beta_{\rm N}, \sqrt{\frac{\mu_1}{\beta_{\rm N}}} , \Delta_2, \mu_1, \mu_2)$.

\noindent
Proof.}  The first claim
 has been indicated by Nishimori \cite{N}.  Here, we prove this claim.
 The identity (\ref{2pointNM})  implies 
 \begin{eqnarray}
\mathbb E  \langle (m^1)^2 \rangle_{\beta} 
&=&\frac{1}{|\Lambda_L|^2} \sum_{i,j\in \Lambda_L}  \mathbb  E \langle \sigma_i \sigma_j  \rangle_{\beta}
= \frac{1}{|\Lambda_L|^2} \sum_{i,j\in \Lambda_L} \mathbb  E \langle \sigma_i \sigma_j  \rangle_{\beta}
   \langle \sigma_i \sigma_j  \rangle_{\beta_{\rm N}} \nonumber \\
  &\leq&   \frac{1}{|\Lambda_L|^2} \sum_{i,j\in \Lambda_L} \mathbb  E |\langle \sigma_i \sigma_j  \rangle_{\beta}||
   \langle \sigma_i \sigma_j  \rangle_{\beta_{\rm N}} | \leq   
   \frac{1}{|\Lambda_L|^2} \sum_{i,j\in \Lambda_L} \mathbb  E   |\langle \sigma_i \sigma_j  \rangle_{\beta_{\rm N}} |\nonumber \\
&\leq&  \frac{1}{|\Lambda_L|^2} \sum_{i,j\in \Lambda_L}\sqrt{ \mathbb  E   \langle \sigma_i \sigma_j  \rangle_{\beta_{\rm N}}}
\leq
 \sqrt{ \frac{1}{|\Lambda_L|^2} \sum_{i,j\in \Lambda_L} \mathbb  E   \langle \sigma_i \sigma_j  \rangle_{\beta_{\rm N}}}\nonumber \\
&\leq &\sqrt{ \mathbb E  \langle (m^1)^2 \rangle_{\beta_{\rm N}} },
\end{eqnarray}
Therefore,  the assumption  implies
\begin{equation}
\lim_{L\to \infty} \mathbb E   \langle (m^1)^2 \rangle_{\beta} =0,
\end{equation}
 also for $(\beta,0, \Delta_2,0, \mu_2)$.
 The identity (\ref{1pointNM}) and the Cauchy-Schwarz inequality  give a bound on
 the  magnetization  for $(\beta, \sqrt{\frac{\mu_1}{\beta_{\rm N}}} , \Delta_2,\mu_1, \mu_2)$ 
\begin{equation}
|{\mathbb E} \langle \sigma_i \rangle_{\beta}|
=|{\mathbb E} \langle \sigma_i \rangle_{\beta}\langle \sigma_i \rangle_{\beta_{\rm N} } |
\leq {\mathbb E}| \langle \sigma_i \rangle_{\beta} ||\langle \sigma_i \rangle_{\beta_{\rm N} } |
\leq{  \mathbb E}| \langle \sigma_i \rangle_{\beta_{\rm N} }|\leq \sqrt{{\mathbb E} \langle \sigma_i \rangle_{\beta_{\rm N} }^2}
  =\sqrt{{\mathbb E} \langle \sigma_i \rangle_{\beta_{\rm N} }},
\end{equation}
for any $i \in \Lambda_L$. This and Jensen's inequality imply
\begin{equation}
|\mathbb E   \langle m^1 \rangle_{\beta}|
\leq \frac{1}{|\Lambda_L|} \sum_{i\in \Lambda_L} 
|{\mathbb E} \langle \sigma_i \rangle_{\beta} |\leq 
 \frac{1}{|\Lambda_L|} \sum_{i\in \Lambda_L} 
\sqrt{{\mathbb E} \langle \sigma_i \rangle_{\beta_{\rm N}}} \leq 
\sqrt{ \frac{1}{|\Lambda_L|} \sum_{i\in \Lambda_L} 
{\mathbb E} \langle \sigma_i \rangle_{\beta_{\rm N}}} =\sqrt{{\mathbb E} \langle m^1 \rangle_{\beta_{\rm N}}}.
\end{equation}
Therefore,  the assumption on the spontaneous magnetization
on  NM implies that
there is no spontaneous magnetization for an arbitrary $\beta >0$
\begin{equation}
\lim_{\mu_1\searrow 0}\lim_{L\to \infty} \mathbb E   \langle m^1 \rangle_{\beta}\leq  \lim_{\mu_1\searrow 0}\lim_{L\to \infty} \mathbb E   \langle m^1 \rangle_{\beta_{\rm N}}
 =0.
\end{equation}
This completes the proof.   
 $\Box$
{\notes
Theorem \ref{cor} is valid also in the model with $Q \supsetneq \{1,2\}$, if the sample expectation of the square of magnetization
and spontaneous magnetization vanish on the NM in this model.\\
Theorem \ref{cor} is proven also in quantum spin glasses, for example in the Edwards-Anderson model with quantum mechanical 
 perturbations \cite{IS}. 
 These results  are well-known general properties of spin glasses, which are consistent with rounding effects
obtained in Ref. \cite{AGL, AW, GAL}. 
Nishimori's  gauge theory is useful  also for disordered  quantum spin systems \cite{MON}. }

 \subsection{Self-averaging of order parameters  on the NM}
 
{\lemma  \label{variance}
For any $p \in Q$ and  for any $\Delta_p>0$, 
the variance of the  ferromagnetic order parameter vanishes
\begin{equation}
\lim_{L\rightarrow \infty} {\mathbb E} \langle( m^p- \langle m^p \rangle)^2 \rangle =0.
\label{variancem1}
\end{equation}
On  the NM,  the variance  of the overlap  vanishes
\begin{equation}
\lim_{L\to\infty}\mathbb E \langle (R_{1,2} ^p - \langle R_{1,2} ^p\rangle )^2 \rangle=0.
\label{varianceR1}
\end{equation}
 in the infinite volume limit.
 \\
Proof.} 
 For any positive integer $k$, any $p \in Q$ and any $ X \in {\cal B}_p$, 
 the summation over truncated correlation functions has an upper bound \cite{C1,I2}
\begin{equation}
\sum_{Y_1,\cdots, Y_k \in {\cal B}_p} ( \mathbb E \langle  \sigma_X;\sigma_{Y_1}; \cdots ; \sigma_{Y_k} \rangle)^2 
\leq k! (\beta \Delta_p)^{-2k}.
\label{Chatterjee-correlation}
 \end{equation}
The inequality (\ref{Chatterjee-correlation}) for $k=1$ implies
\begin{equation}
\sum_{Y \in {\cal B}_p} [ \mathbb E( \langle  \sigma_X\sigma_{Y} \rangle-
\langle  \sigma_X\rangle \langle \sigma_{Y} \rangle)]^2 \leq (\beta \Delta_p)^{-2}.
\end{equation}
This and the Cauchy-Schwarz inequality imply
\begin{equation}
\sum_{Y \in {\cal B}_p}  |\mathbb E( \langle  \sigma_X\sigma_{Y} \rangle-
\langle  \sigma_X\rangle \langle \sigma_{Y} \rangle )|
\leq 
\sqrt{\sum_{Y \in {\cal B}_p} [ \mathbb E( \langle  \sigma_X\sigma_{Y} \rangle-
\langle  \sigma_X\rangle \langle \sigma_{Y} \rangle)]^2 
\sum_{Y \in {\cal B}_p} 1^2}\leq  \frac{ \sqrt{| {\cal B}_p|}}{\beta\Delta_p}.
\label{k1}
\end{equation}
From the above inequality, the variance of the ferromagnetic order parameter is obtained
 \begin{equation}
\lim_{L\to \infty} {\mathbb E} \langle( m^p- \langle m^p \rangle)^2 \rangle=
\lim_{L\to \infty} \frac{1}{|{\cal B}_p|^2} \sum_{X,Y \in {\cal B}_p}  \mathbb E( \langle  \sigma_X\sigma_{Y} \rangle-
\langle  \sigma_X\rangle \langle \sigma_{Y} \rangle)
 \leq \lim_{L\to\infty} \frac{1}{\beta\Delta_p \sqrt{| {\cal B}_p|}}=0,
 \end{equation}
 since $|{\cal B}_p|$ is proportional to $|\Lambda_L|$. This  gives the first identity (\ref{variancem1}).
The inequality (\ref{Chatterjee-correlation}) for $k=3$ for the summation over 
$Y_1=X$ and $Y_2=Y_3=Y$ implies 
\begin{equation}
\sum_{Y \in {\cal B}_p} [ \mathbb E( \langle  \sigma_X \sigma_{Y} \rangle^2-4
\langle  \sigma_X\rangle \langle \sigma_{Y} \rangle \langle  \sigma_X \sigma_{Y} \rangle+3
\langle  \sigma_X\rangle^2 \langle \sigma_{Y} \rangle^2)]^2 \leq \frac{3}{2}(\beta \Delta_p)^{-6}.
\end{equation}
Therefore, 
 \begin{equation}
\sum_{Y \in {\cal B}_p}|  \mathbb E( \langle  \sigma_X \sigma_{Y} \rangle^2-4
\langle  \sigma_X\rangle \langle \sigma_{Y} \rangle \langle  \sigma_X \sigma_{Y} \rangle
+3\langle  \sigma_X\rangle^2 \langle \sigma_{Y} \rangle^2) |
\leq \frac{1}{\beta^3 \Delta_p^{3}} \sqrt{\frac{3 |{\cal B}_p|}{2}}.
\end{equation}
On the NM, this inequality is represented as
\begin{equation}
\sum_{Y \in {\cal B}_p}|  \mathbb E( \langle  \sigma_X \sigma_{Y} \rangle-4
\langle  \sigma_X\rangle \langle \sigma_{Y} \rangle 
+3\langle  \sigma_X\rangle^2 \langle \sigma_{Y} \rangle^2) |
\leq  \frac{1}{\beta_{\rm N}^3 \Delta_p^{3}} \sqrt{\frac{3 |{\cal B}_p|}{2}}.
\end{equation}
This and the inequality (\ref{k1}) imply
\begin{eqnarray}
&&
\sum_{Y \in {\cal B}_p} | \mathbb E(\langle  \sigma_X\rangle^2 \langle \sigma_{Y} \rangle^2
-\langle  \sigma_X\rangle \langle \sigma_{Y} \rangle ) |
 \nonumber \\
&&
=  \frac{1}{3}\sum_{Y \in {\cal B}_p}  
|\mathbb E( - \langle  \sigma_X \sigma_{Y} \rangle+ \langle  \sigma_X\rangle \langle \sigma_{Y} \rangle+ \langle  \sigma_X \sigma_{Y} \rangle
-4\langle  \sigma_X\rangle \langle \sigma_{Y} \rangle 
+3\langle  \sigma_X\rangle^2 \langle \sigma_{Y} \rangle^2) | \nonumber \\ 
&&
\leq
 \frac{1}{3} \sum_{Y \in {\cal B}_p} |\mathbb E(  \langle  \sigma_X \sigma_{Y} \rangle- \langle  \sigma_X\rangle \langle \sigma_{Y} \rangle )|+
 \frac{1}{3} \sum_{Y \in {\cal B}_p}  | {\mathbb E} ( \langle  \sigma_X \sigma_{Y} \rangle
-4\langle  \sigma_X\rangle \langle \sigma_{Y} \rangle +3\langle  \sigma_X\rangle^2 \langle \sigma_{Y} \rangle^2) | \nonumber \\
&&
\leq\Big(\frac{1}{3}+\frac{1}{\sqrt{6 }\beta_{\rm N}^2 \Delta_p^{2}} \Big) \frac{ \sqrt{| {\cal B}_p|}}{\beta_{\rm N}\Delta_p},
\label{Ch}
\end{eqnarray}
and 
\begin{eqnarray}
&&\sum_{Y \in {\cal B}_p}  \mathbb E(\langle  \sigma_X \sigma_{Y} \rangle^2-
\langle  \sigma_X\rangle^2 \langle \sigma_{Y} \rangle^2
)\nonumber \\
&&=\sum_{Y \in {\cal B}_p}  \mathbb E(\langle  \sigma_X \sigma_{Y} \rangle- \langle  \sigma_X\rangle \langle \sigma_{Y} \rangle
+ \langle  \sigma_X\rangle \langle \sigma_{Y} \rangle-
\langle  \sigma_X\rangle^2 \langle \sigma_{Y} \rangle^2
) \nonumber  \\
&& \leq\sum_{Y \in {\cal B}_p} | \mathbb E(\langle  \sigma_X \sigma_{Y} \rangle- \langle  \sigma_X\rangle \langle \sigma_{Y} \rangle ) | 
+\sum_{Y \in {\cal B}_p} | \mathbb E( \langle  \sigma_X\rangle \langle \sigma_{Y} \rangle-
\langle  \sigma_X\rangle^2 \langle \sigma_{Y} \rangle^2
)| \nonumber  \\
&& \leq
\Big(\frac{4}{3}+\frac{1}{\sqrt{6 }\beta_{\rm N}^2 \Delta_p^{2}} \Big)\frac{ \sqrt{| {\cal B}_p|}}{\beta_{\rm N}\Delta_p}.
\end{eqnarray}
The following variance of overlap  is bounded by
\begin{eqnarray}
\mathbb E \langle (R_{1,2} ^p - \langle R_{1,2} ^p\rangle )^2 \rangle&=& \frac{1}{| {\cal B}_p|^2}\sum_{X,Y \in {\cal B}_p}  \mathbb E(\langle  \sigma_X \sigma_{Y} \rangle^2-
\langle  \sigma_X\rangle^2 \langle \sigma_{Y} \rangle^2
)  \\
&
\leq&  \Big(\frac{4}{3}+\frac{1}{\sqrt{6 }\beta_{\rm N}^2 \Delta_p^{2}} \Big)\frac{ 1}{\beta_{\rm N}\Delta_p\sqrt{| {\cal B}_p|}}.
\end{eqnarray}
Since $| {\cal B}_p|$ is proportional to $L^d$ for any $p\in Q$,  the variance of the overlap vanishes on the NM in the infinite-volume limit 
\begin{equation}
\lim_{L\to\infty}\mathbb E \langle (R_{1,2} ^p - \langle R_{1,2} ^p\rangle )^2 \rangle=0.
\end{equation}
This completes the proof. $\Box$ 
\\

The following lemma has been given by  Aizenman-Contucci  and Ghirlanda-Guerra independently
  \cite{AC,GG}.
  {\lemma \label{ACGG}(Aizenman-Contucci-Ghirlanda-Guerra identities)
Let  $f: \Sigma_L^n \to  {\mathbb R}$ be a bounded function, where $\Sigma_L := \{-1,1 \}^{\Lambda_L}$ 
is a set of  all spin configurations. 
The Aizenman-Contucci-Ghirlanda-Guerra (ACGG) identities 
give
\begin{equation}\lim_{L \to \infty} ( \mathbb{E} \langle f R_{1, n+1}^p \rangle - \frac{1}{n} \mathbb{E} \langle f \rangle \mathbb{E} \langle R_{1,2}^p \rangle - \frac{1}{n} \sum_{a=2}^n \mathbb{E} \langle f R_{1,a} ^p\rangle )= 0,
\label{ACGGI}
\end{equation}
 for any bounded  $f$, for any $p\in Q$  and for almost all $\Delta_p >0$.\\
 Proof.}  
 The proof is  given in several literatures 
  \cite{AC,C1,C2,CG,CG2,CG3,CGN,GG,I2,I,T}.  
   The ACGG  identities are proven on the basis of  the self-averaging property of $\psi_L(\beta, \bm J)$ and
   the convexity of  
   $$p(\beta, \bm \Delta, \bm \mu):= \lim_{L\to \infty}p_L(\beta, \bm \Delta, \bm \mu),$$ with respect to $\Delta_p$. 
 $\Box$
{\theorem  \label{MT}  For any $p \in Q$, for almost all  $\mu_p >0$, 
the variance of the  ferromagnetic order parameter vanishes
\begin{equation}
\lim_{L\rightarrow \infty} {\mathbb E} \langle( m^p-{\mathbb E} \langle m^p \rangle)^2 \rangle =0,
\end{equation}
 in the infinite volume limit.
 For any $p \in Q$,  for almost all $\Delta_p>0$ on the NM, 
the variance of the overlap vanishes 
\begin{equation}
\lim_{L\rightarrow \infty} {\mathbb E} \langle( R_{1,2} ^p-{\mathbb E} \langle R_{1,2}^p \rangle)^2 \rangle =0,
\end{equation} 
in the infinite volume limit.
\\

\noindent
Proof.} The self-averaging property of the magnetization $\langle m^p \rangle$ 
for almost all $\mu_p > 0$  is proven as in the same method to prove the ACGG identities.
The proof  is given  on the basis of  the self-averaging property of $\psi_L(\beta, \bm J)$  
and convexity of $
p(\beta, \bm \Delta, \bm \mu)$ with respect to $\mu_p$  \cite{CGN,I2}. 
\\
Here, we  prove  the self-averaging property
of  the overlap  on NM.
\\For  $n=2, f = R_{1,2}^p$, ACGG identities (\ref{ACGGI}) give
\begin{eqnarray}
\lim_{L \to \infty} [ \mathbb{E} \langle R_{1,2}^p R_{1,3}^p \rangle - \frac{1}{2} ( \mathbb{E} \langle R_{1,2}^p \rangle )^2  - \frac{1}{2}\mathbb{E} \langle( R_{1,2}^p)^2 \rangle ]= 0. \label{k2}
\end{eqnarray}
For $n=3, f = R_{2,3}$,  
\begin{eqnarray}
\lim_{L \to \infty} [ \mathbb{E} \langle R_{2,3}^p R_{1,4}^p \rangle - \frac{1}{3}\mathbb{E} \langle R_{2,3}^p \rangle\mathbb{E} \langle R_{1,2} ^p\rangle  
- \frac{1}{3} \mathbb{E} \langle R_{2,3}^p R_{1,2}^p \rangle   - \frac{1}{3} \mathbb{E}\langle R_{2,3}^p R_{1,3}^p \rangle ]= 0.
\end{eqnarray}
The replica symmetry implies $  \langle R_{2,3} ^p \rangle= \langle R_{1,2} ^p \rangle$, 
\begin{eqnarray*}
\langle R_{2,3}^p R_{1,2}^p \rangle =
   \langle R_{2,3} ^p R_{1,3}^p \rangle =  \langle R_{1,3}^p R_{1,2}^p \rangle,
\end{eqnarray*}
 and $\langle R_{2,3}^p R_{1,4}^p \rangle = \langle R_{1,2} ^p \rangle^2$, then we have
\begin{eqnarray}
\lim_{L \to \infty} [ \mathbb{E} \langle R_{1,2}^p \rangle^2 - \frac{1}{3}(\mathbb{E} \langle R_{1,2}^p \rangle)^2 - \frac{2}{3} \mathbb{E} \langle R_{1,3}^p R_{1,2}^p \rangle   ]= 0.
 \label{k3}
\end{eqnarray}
Substitute the identity (\ref{k2}) into (\ref{k3}), then we have
\begin{eqnarray*}
\lim_{L \to \infty} [ \mathbb{E} \langle R_{1,2}^p \rangle^2 - \frac{2}{3} ( \mathbb{E} \langle R_{1,2}^p \rangle )^2  - \frac{1}{3} \mathbb{E}\langle (R_{1,2}^p)^2 \rangle ]= 0.
\end{eqnarray*}
 For two deviations
$R_{1,2}^p -{\mathbb E}\langle  R_{1,2} ^p \rangle $ and $R_{1,2}^p -\langle  R_{1,2}^p  \rangle $,
there are relations between two  variances
\begin{equation}
2\lim_{L \rightarrow \infty} {\mathbb E}\langle (R_{1,2}^p -{\mathbb E}\langle  R_{1,2} ^p\rangle )^2 \rangle
=3\lim_{L \rightarrow \infty} {\mathbb E}\langle (R_{1,2}^p -\langle  R_{1,2} ^p\rangle )^2 \rangle,
\label{D-d}
\end{equation}
This and the identity (\ref{varianceR1}) imply
\begin{equation}
\lim_{L \rightarrow \infty} {\mathbb E}\langle (R_{1,2}^p -{\mathbb E}\langle  R_{1,2} ^p\rangle )^2 \rangle=0,
\end{equation}
for any $p\in Q$ and for almost all $\Delta_p >0$ on the NM.
This has proven Theorem \ref{MT}. $\Box$

\paragraph{Example: } Consider a model defined by $Q:=\{1,2\}$, ${\cal B}_1 := \Lambda_L$ and  nearest neighbor  bonds
${\cal B}_2:= \{\{i,j \} | i,j\in \Lambda_L,  |i-j|=1,\}$. 
This model is the random field  Edwards-Anderson model.
Theorem \ref{MT} is still valid in a $\mathbb Z_2$-symmetric limit  $ \Delta_1 \to 0$  on the NM defined by  $\beta_{\rm N} = \mu_1/\Delta_1^2
 = \mu_2/\Delta_2^2$, even though spontaneous $\mathbb Z_2$ symmetry breaking appears.

\section{Mean field mixed $p$-spin glasses
} 
\subsection{Definitions} 
Let $L$ be a positive integer and define  $\Lambda_L:= [1,L] \cap \mathbb Z$.  
A collection 
$${\cal B}_p:= \{ X \subset \Lambda_L  | |X|=p  \},
$$ consists of all subsets of $\Lambda_L$ with cardinality $p$. 
Mean field mixed $p$-spin glass model is defined by
\begin{equation}
H(\sigma, \bm J):=-\sum_{p \in Q} 
\sum_{X \in {\cal B}_p}J^p_X
\sigma_X
\end{equation}
The probability density function of $p$-spin interaction is
defined by
\begin{equation}
P_p(J^p_X
):=\sqrt{\frac{L^{p-1}}{2\pi \Delta_p^2} }\exp \Big[- \frac{L^{p-1}}{2\Delta_p^2}(J^p_X
 -L^{1-p}\mu_p) ^2\Big]. 
 \label{MFdistributionJ}
\end{equation}
If $\Delta_p=0=\mu_p$ for all odd $p \in Q$, the model has $\mathbb Z_2$-symmetry, 
as in the short-range model.
The coupling constant $J_X^p$ is represented in terms of 
standard Gaussian random variables $(g_X^p)_{X \in {\cal B}_p}$
\begin{equation}
J_X^p = L^{(1-p)/2} \Delta_p g_X^p + L^{1-p} \mu_p. 
\end{equation}
Also in the mean field models, NM is defined by $\beta_{\rm N} = \mu_p/\Delta_p^2$ for all $p\in Q$ as in short-range models. 
 Lemma \ref{NM}, \ref{variance}, \ref{ACGG}, 
Theorem \ref{cor} and Theorem \ref{MT} are all valid also in the mean field mixed $p$-spin glass model. 
The mean field model with $Q=\{2\}$ and coupling constants $(\beta, 0, \Delta_2, 0, \mu_2)$
is the SK model and that with  $Q=\{1,2\}$ and coupling constants $(\beta, \Delta_1, \Delta_2, \mu_1, \mu_2)$
is the SK model with a Gaussian random field. 
In the SK model,  the assumptions are satisfied in Theorem \ref{cor},  
since  the Sherrington-Kirkpatrick (SK) solution is valid on the NM  and the
spontaneous magnetization vanishes  
on the NM \cite{C-WK1,C-WK2}.

\section{Conclusions and discussions}
In the present paper, we provide two main theorems for mixed $p$-spin glass models both in finite and infinite dimensions. 

\begin{figure}[H]
\begin{center}
\includegraphics{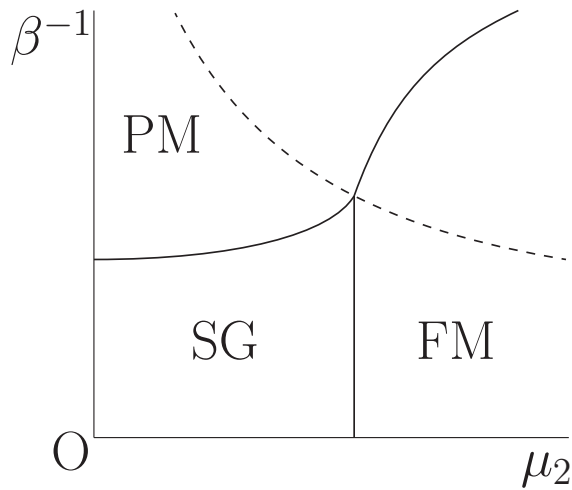}
\end{center}
\caption{
Phase diagram in the Edwards-Anderson model for $p=2$. 
The ordinate is the temperature $\beta^{-1}$, and the abscissa is the ferromagnetic 
exchange interaction $\mu_2$ between two spins. 
The solid lines denote the phase boundaries among the ferromagnetic (FM), paramagnetic (PM) and spin glass (SG) phases. 
The broken line denotes the Nishimori line. Theorem \ref{cor} implies that sample expectation of 
the spontaneous magnetization does not
exist. This fact guarantees that
 the phase boundary between spin glass and ferromagnetic phases is conjectured to be vertical \cite{N0}.  
Reproduced with permission from \cite{N0}. 
}
\label{figNL}
\end{figure}

 Theorem \ref{cor} characterizes the nature of phase diagram 
 in the Edwards-Anderson model.
First, we remark the definitions of the paramagnetic, spin glass and ferromagnetic phases in terms of two order parameters 
 \begin{equation}
m_+:= \lim_{\mu_1 \searrow 0} \lim_{L\to\infty}\mathbb E \langle m^1 \rangle_\beta, \ \ \ 
q_+:=\lim_{\mu_1 \searrow 0} \lim_{L\to\infty}\mathbb E \langle R_{1,2} \rangle_\beta.
 \end{equation} 
 The paramagnetic, spin glass and ferromagnetic phases are defined by $m_+ =0, q_+=0$, $m_+=0, q_+ > 0$ and $m_+>0,q_+>0$, respectively.  
The first claim in Theorem \ref{cor} 
concludes that
 the sample expectation of the square of the ferromagnetic
 magnetization vanishes 
 \begin{equation}\lim_{L\to\infty}\mathbb E \langle (m^1)^2 \rangle_\beta=0, \end{equation}
   both in paramagnetic and spin glass phases 
 in the Edwards-Anderson model for $(\beta, \Delta_1,\Delta_2, \mu_1,\mu_2)=(\beta,0,\Delta_2,0,\mu_2)$.
 The first claim
 can be shown by the correlation identity (\ref{2pointNM}) already obtained by Nishimori in the original paper \cite{N0,N}. 
  The second claim in Theorem \ref{cor} concludes 
  also that  spontaneous magnetization vanishes 
  \begin{equation}
m_+=\lim_{\mu_1 \searrow 0} \lim_{L\to\infty}\mathbb E \langle m^1 \rangle_\beta=0,
 \end{equation}
 for any $(\beta, \Delta_2, \mu_2)$, if $m_+=0$ for the zero field limit with $\Delta_1=\sqrt{\mu_1 /\beta_{\rm N} }$
 on the NM  
 for $(\beta_{\rm N}, \Delta_2, \mu_2)$ defined by $\beta_{\rm N}:= \mu_2/\Delta_2^2$.
 To show the second claim for no spontaneous magnetization, the identity (\ref{1pointNM}) in 
 the mixed $p$-spin model with $p=1$ and $p=2$ is utilized. It is stressed that
 these useful identities given by Nishimori's gauge theory are valid 
 in the mixed $p$-spin glass models even under $\mathbb Z_2$-symmetry breaking field.  
 This result constrains the phase diagram of the Edwards-Anderson model.
 The ferromagnetic phase defined by $m_+>0, q_+>0$ does not exist for 
any $(\beta, \Delta_2, \mu_2)$, if the point $(\beta_{\rm N}, \Delta_2, \mu_2)$ is in 
the paramagnetic phase defined by $m_+=0=q_+$.
This is consistent with the conjecture \cite{N0}
that the phase boundary between spin glass and ferromagnetic phases depends only on $\mu_2$ as depicted 
in Fig. \ref{figNL}.
Theorem \ref{MT} concludes the absence of RSB on the Nishimori line in a rigorous manner.
Identities given by Nishimori's gauge theory combined with
the ACGG identities \cite{AC,C1,C2,CG,CG2,CG3,CGN,GG,I2,I,T} enable us to
prove that variances of the magnetization and the spin overlap vanish on the Nishimori line. 
Nishimori and Sherrington obtained the same result \cite {NS} by showing
that the distribution function of the spin overlap is identical to that of the magnetization on the Nishimori line. 
Their result for the overlap depends on the assumption that the distribution of the magnetization is concentrated at 
a single value on the Nishimori line.  We have proven this concentration property of the magnetization and the spin overlap
 in the same way to obtain the ACGG identities.

\paragraph{Acknowledgments}
C.I. is supported by JSPS (21K03393).\\
\\
There is no conflict of interest. All data are provided in full in this paper.



\begin{thebibliography}{99}

\bibitem{AC} Aizenman, M. and Contucci, P. :{On the stability of quenched state in mean-field spin glass models}. J. Stat. Phys. \textbf{92}, 765-783 (1997)

\bibitem{AGL} Aizenman, M., Greenblatt, R. L. and Lebowitz, J. L. :{Proof of rounding by quenched disorder of first order transitions in low-dimensional quantum systems}. J. Math. Phys. {\textbf{53}}, 023301 (2012)

\bibitem{AW} Aizenman, M. and Wehr, J. :{Rounding effects of quenched randomness on first-order phase transitions}.  Commun. Math. Phys. \textbf{130}, 489-528 (1990) 

\bibitem{ABBB} Agliari, E., Barra, A., Burioni, R. and Di Biasio, A. :
{Notes on the p-spin glass studied via Hamilton-Jacobi and smooth-cavity techniques}. 
J. Math. Phys. \textbf{53}, 063304 (2012)

\bibitem{AFM} Agliari, E., Fachechi, A. and Marullo, C. :
{Nonlinear PDEs approach to statistical mechanics of dense associative memories}. 
J. Math. Phys. \textbf{63}, 103304 (2022)

\bibitem{AA} Albanese, L. and Alessandrelli, A. :
{Rigorous approaches for spin glass and Gaussian spin glass with P-wise interactions}. 
preprint, arXiv:2111.12569 (2021)

\bibitem{ACCM} Alberici, D., Camilli, F., Contucci, P. and Mingione, E. :
{The multi-species mean-field spin-glass on the Nishimori line}. 
J. Stat. Phys. \textbf{182}, 2 (2021)


\bibitem{BBG} Barra, A., Di Biasio, A. and Guerra, F. :
{Replica symmetry breaking in mean-field spin glasses through the Hamiltonian-Jacobi technique}. 
J. Stat. Mech. P09006 (2010)

\bibitem{BFT} Barra, A., Dal Ferraro, G. and Tantari, D. :
{Mean field spin glasses treated with PDE techniques}. 
Eur. Phys. J. B \textbf{86}, 332 (2013)

\bibitem{KXZ} 
Barbier, J., Dia, M., Macris, N., Krzakala, F., Lesieur, T. and Zdeborov\'a, L. : 
{Mutual information for symmetric rank-one matrix estimation: A proof of the replica formula}. 
Advances in Neural Information Processing Systems \textbf{29}, 424-432 (2016)

\bibitem{BDMKZ} Barbier, J., Dia, M., Macris, N., Krzakala, F. and Zdeborov\'a, L. :
{Rank-one matrix estimation: analysis of algorithmic and information theoretic limits by the spatial coupling method}. preprint, arXiv:1812.02537 (2018)

\bibitem{C1} Chatterjee, S. :{The Ghirlanda-Guerra identities without averaging}. preprint, arXiv:0911.4520 (2009).

\bibitem{C2} Chatterjee, S. :{Absence of replica-symmetry breaking in the random field Ising model}. 
Commun. Math .Phys. \textbf{337}, 93-102 (2015)





\bibitem{C-WK1} Chen, W-K. :{On the mixed even-spin  Sherrington-Kirkpatrick model with ferromagnetic interaction}. 
Ann. Henri Poincar\'{e}  Probab. Stat. \textbf{50}, 63-83 (2014) 

\bibitem{C-WK2} Chen, W-K. :{On the Almeida-Thouless transition line in the Sherrington-Kirkpatrick model with centered Gaussian external field}. 
Electron. Commun. Probab. {\bf 26}, 65, 1-9 (2021)

\bibitem{CG} Contucci, P. and Giardin\`a, C. :{Spin-glass stochastic stability: A rigorous proof}. 
Ann. Henri Poincare, \textbf{6}, 915-923 (2005)  

\bibitem{CG2} Contucci, P. and Giardin\`a, C. :{The Ghirlanda-Guerra identities}. 
J. Stat. Phys. \textbf{126}, 917-931 (2007)

\bibitem{CG3} Contucci, P. and Giardin\`a, C. :{Perspectives on spin glasses.} Cambridge university press, 2012.

%
\bibitem{CGN} Contucci, P., Giardin\`a, C. and Nishimori, H. :{Spin glass identities and the Nishimori line}. 
In Spin Glasses: Statics and Dynamics 103-121 Springer, (2009).

\bibitem{CGP} Contucci, P., Giardin\`a, C. and Pul\'e, J. :{The infinite volume limit for finite dimensional classical and quantum disordered systems}. Rev.  Math. Phys. \textbf{16}, 629-638 (2004)

\bibitem{CL} Contucci, P. and Lebowitz, J. L. :{Correlation inequalities for quantum spin systems with quenched centered disorder}. 
J. Math. Phys. \textbf{51}, 023302-1 -6 (2010)


\bibitem{AT}de Almeida, J. R. L. and Thouless, D. J. :{Stability of the Sherrington-Kirkpatrick solution of spin glass model}. 
J. Phys. A :Math. Gen. \textbf{11}, 983-990 (1978)





\bibitem{EA}  Edwards, S. F. and Anderson, P. W. :{Theory of spin glasses}. J. Phys. F: Metal Phys. \textbf{5}, 965-974 (1975)




\bibitem{G} Guerra, F. :{Sum rules for the free energy in the mean field spin glass model}. 
Fields Institute Communications \textbf{30}, 161-170 (2001)

\bibitem{G1} Guerra, F. :{Broken replica symmetry bounds in the mean field spin glass model}. 
Commun. Math. Phys. {\bf 233}, 1-12 (2003)

\bibitem{GG} Ghirlanda, S. and Guerra, F. :{General properties of overlap probability distributions in disordered spin systems. Towards Parisi ultrametricity}. 
J. Phys. A:Math. Gen. \textbf{31}, 9149-9155 (1998)



\bibitem{GAL} Greenblatt, R. L., Aizenman, M. and Lebowitz, J. L. :{Rounding first order transitions in low-dimensional quantum systems with quenched disorder}. 
Phys. Rev. Lett. {\textbf{103}}, 197201 (2009)

 



%





\bibitem{I2} Itoi, C. :{Universal nature of replica-symmetry breaking in disordered quantum  systems}. 
J. Stat. Phys. {\textbf{167}}, 1262-1279  (2017)
 
\bibitem{I} Itoi, C. :{Self-averaging of perturbation Hamiltonian density in perturbed spin systems}. 
J. Stat. Phys. \textbf{177}, 1063-1076 (2019) 

\bibitem{IS} Itoi, C. and Sakamoto, Y. :{Boundedness of Susceptibility in Spin Glass Transition 
of Transverse Field Mixed $p$-spin Glass Models}. 
J. Phys. Soc. Jpn. \textbf{92}, 064004 (2023) 




 










\bibitem{MNC} Morita, S., Nishimori, H. and Contucci, P. :{Griffiths inequalities for the Gaussian spin glass}. 
J. Phys. A: Math. Gen. {\bf 37}, L203-211 (2004)

\bibitem{MON} Morita,S., Ozeki, Y. and Nishimori, H. :{Gauge theory for quantum spin glasses}. 
J. Phys. Soc. Jpn. {\bf 75}, 014001, 1-7 (2006)


\bibitem{N0} Nishimori, H. :{Statistical physics of spin glasses and information processing an introduction}. 
Oxford university press. (2001)

\bibitem{N} Nishimori, H. :{Internal energy, specific heat and correlation function of the bond-random Ising model}. 
Prog. Theor. Phys. {\bf 66}, 1169-1181 (1981)

 
\bibitem{NS} Nishimori, H. and Sherrington, D. :{Absence of replica-symmetry breaking in a region of the phase diagram of the Ising spin glass}. 
AIP conference proceedings {\bf 553}, 67 (2001)
  
\bibitem{OO} Okuyama, M. and Ohzeki, M. :{Gibbs-Bogoliubov inequality on Nishimori line}. preprint, arXiv:2208.12311 (2022) 



\bibitem{Pr0} Parisi, G. :{Infinite number of order parameters for spin glasses}. 
Phys. Rev. Lett. \textbf{43}, 1754-1756 (1979).


\bibitem{Pr} Parisi, G. :{A sequence of approximate solutions to the S-K model for spin glasses}. 
J. Phys. A: Math. Gen. \textbf{13}, L115-L121 (1980).



\bibitem{SK} Sherrington, D. and Kirkpatrick, S. :{Solvable model of spin glass}. Phys. Rev. Lett. \textbf{35}, 1792-1796 (1975) 


\bibitem{T2}  Talagrand, M. :{The Parisi formula}. Ann. Math. \textbf{163}, 221-263 (2006)

\bibitem{T} Talagrand, M. :{Mean field models for spin glasses}. Springer, Berlin (2011)
	 
\bibitem{To} Toninelli, F. :{About the Almeida-Thouless transition line in the Sherrington-Kirkpatrick mean field spin glass model}. 
Euro. Phys. Lett. {\bf 60}, 764-767 (2002)






 



 
 

\end{thebibliography}
\end{document}